# Cercignani's conjecture is true for Smoluchowski coagulation equation

Mingliang Xie

State Key Laboratory of Coal Combustion, Huazhong University of Science and Technology, Wuhan 430074, China

Email: mlxie@mail.hust.edu.cn

Abstract: In the present study, the information entropy for Smoluchowski coagulation equation is proposed based on the statistical physics. And the normalized particle size distribution is a lognormal function at equilibrium from the principle of maximum entropy and moment constraint. The parameters in the particle size distribution are determined as simple constants, the result reveals that the assumption that algebraic mean volume be unit in self-preserving hypothesis is reasonable. Based on the present definition of the information entropy, the Cercignani's conjecture holds naturally for Smoluchowski coagulation equation. Together with the fact that the conjecture is also true for Boltzmann equation, Cercignani's conjecture holds for any two-body collision systems, which will benefit the understanding of Brownian motion and molecule kinematic theory, such as the stability of the dissipative system, and the mathematical theory of convergence to thermodynamic equilibrium.

Keywords: particle size distribution; population balance equation; information entropy; moment method; Brownian motion; convergence

**Introduction**

Population balance equation (PBE) are general mathematical framework for modeling of particulate system. In the framework of mono-variants internal coordinate and time for each particle, the PBE characterized as Smoluchowski coagulation equation (SCE), which takes form (Friedlander, 2000):

$$\frac{\partial n(v,t)}{\partial t} = \frac{1}{2}\int_0^v \beta(v_1, v-v_1)n(v_1,t)n(v-v_1)dv_1 - \int_0^\infty \beta(v_1,v)n(v_1,t)n(v,t)dv_1 \qquad (1)$$

In which $n(v,t)dv$ is the number density of particles per unit spatial volume with particle volume from $v$ to $v+dv$ at time $t$, and $\beta$ is the collision frequency function of coagulation. In the free molecule regime, the collision frequency function for Brownian coagulation is

$$\beta = B_1 \left(\frac{1}{v} + \frac{1}{v_1}\right)^{1/2} \left(v^{1/3} + v_1^{1/3}\right)^2 \qquad (2)$$

Where the constant $B_1 = (3/4\pi)^{1/6}(6k_BT/\rho_p)^{1/2}$, $k_B$ is the Boltzmann constant; $T$ is the temperature; and $\rho_p$ is the particle density. In the continuum regime, the corresponding collision frequency function is

$$\beta = B_2 \left(\frac{1}{v^{1/3}} + \frac{1}{v_1^{1/3}}\right)\left(v^{1/3} + v_1^{1/3}\right) \qquad (3)$$

Where the constant $B_2 = 3k_BT/2\mu$, and $\mu$ is the gas viscosity.

For the nonlinear partial integrodifferential structure of SCE, only a limited number of known analytical solution exist for simple coagulation kernel (White, 1980; Leyvraz, 2003; Blum, 2006; Niethammer, 2019). If the collision frequency function is a homogeneous function of its arguments, the SCE can be converted into an ordinary integrodifferential equation by a similarity transformation (Friedlander & Wang, 1966). The previous studies indicate that the particle size distribution (PSD) of an aged coagulating system approaches a universal asymptotic form called the self-preserving PSD (Vicsek & Family, 1984; Van Dongen & Ernst, 1985, 1988; Dekkers & Friedlander, 2002; Leyvraz, 2006). If the self-preserving hypothesis can be firmly established, the tedious job of the determination of PSD can be reduced to the determination of only a few parameters for aged system (Pratsinis, 1988; Van Dongen, 1989; Park et al., 1999; Otto et al., 1999; Canizo & Throm, 2021). Although the self-preserving hypothesis has been verified by some experiments and numerical methods (Vemury et al., 1994; Vemury & Pratsinis, 1995; Dekkers et al., 2002), it is unfortunately that the algebraic mean volume of PSD is assumed to be unit in the derivation. This assumption turns the similarity transformation into a simple separated variable method, which makes the self-preserving hypothesis subject to great constraints in theoretically.

Because of the relative simplicity of implementation and low computational costs, the moment method has become a powerful tool for investigating the SCE (Wooldridge, 1998; Aldous, 1999; Rigopoulos, 2000; Roth, 2007; Liao & Lucas, 2010). Recently, Yu et al. (2008) proposed a moment-based approach called the Taylor-series expansion method of moment (TEMOM) to analyze the evolution of particles number density due to Brownian motion. The main idea of the TEMOM is that the nonlinear collision kernel is approximated by a Taylor-series polynomial, and the higher and fractional order moments based on PSD are approximated by the first three integer order moments, and the original TEMOM model achieves self-closure. In addition, the derivation of the TEMOM is completely based on a mathematical method, and no artificial assumption is introduced. The estimated truncation error of TEMOM model for Brownian coagulation has been examined (Xie, 2016), and the particle dimensionless moments corresponding to the deviation of PSD tends to a constant (Xie & Wang, 2013), which is consistent with the self-preserving hypothesis. This method provides a new way for analyzing the coagulation problem theoretically. The moment method can only get finite moments, but the PSD is equivalence with the infinite moments. How to get the PSD from the finite moments is still a hot issue and challenge in science and engineering (John et al., 2007). In addition, the expansion point is usually selected as the algebraic mean volume, whose value increases with time according to the asymptotic solution (Xie & Wang, 2013). Whether there is an equilibrium state cannot be determined by the TEMOM itself.

According to the rule of statistical physics, Shannon information entropy is interpreted as a state function of a thermodynamic system and is proportional to the total number of particles (Jaynes, 1957). The reduction in the total number of particles corresponds to a reduction in the information entropy of a particle system, which must be accompanied by a change of system energy in a dissipative system in accordance with the second low of thermodynamics. Brownian coagulation can be considered as a perfectly inelastic collision process, in which the colliding particles stick together and the maximum amount of kinetic energy of the system is lost. Then the thermodynamic constraint of Brownian coagulation can be proposed based on the binary perfectly inelastic collision theory and principle of maximum entropy (Xie, 2017; Xie & Yu, 2018), which can be regarded as an adjoint equation of the SCE. The thermodynamic constraint can be used to explain the meaning of Eotvos constant in statistical mechanics (Palit, 1956), and its value can be quantitatively calculated by the equality of the constraint at equilibrium. The constraint also gives the expression of the critical time to reach the thermodynamic equilibrium, which can be used to determine whether the PSD reaches the self-preserving form. In addition, the critical time is proportional to temperature, which can explain the long-time existence of urban fine particulate matter in northern China (Zhang et al., 2015), and the structure of macroscopic agglomerates (Blum, 2004). However, many other physical quantities are introduced into the thermodynamic analysis, such as specific surface free energy, internal energy and chemical potential, etc. These physical quantities are not strongly related to the SCE itself, and makes the analysis more complicated and quasi-empirical.

In this study, the information entropy is defined based on the SCE itself. According to the principle of maximum entropy and moment method, it is tried to give the PSD of SCE at equilibrium, and determine the expression for the algebraic mean volume and standard deviation of PSD.

**The TEMOM and its asymptotic solution for Brownian coagulation**

In the moment method, the k-th order moment $M_k$ of PSD is defined as

$$M_k = \int_0^\infty v^k n(v,t) dv \tag{4}$$

And the evolution of particle moment based on the SCE becomes

$$\frac{dM_k}{dt} = \frac{1}{2} \int_0^\infty \int_0^\infty [(v+v_1)^k - v^k - v_1^k] \beta(v,v_1) n(v,t) n(v_1,t) dv_1 \, dv \tag{5}$$

The minimum set of moments required to close the particle moment equation is the first three, $M_0$, $M_1$ and $M_2$. The zeroth order moment represents the total particle number concentration; the first order moment is proportional to the total particle mass concentration, which remains constant due to the rigorous mass conservation requirement; and the second order moment describes the dispersion of PSD. Based on the detailed derivations of TEMOM model (Yu et al., 2018), the resulting closed-form equations for the collision frequency function in free molecule regime can be rearranged as (Xie & Wang, 2013)

$$\frac{dM_0}{dt} = \frac{\sqrt{2} B_1 (65 M_C^2 - 1210 M_C - 9223) M_0^2}{5184} \left(\frac{M_1}{M_0}\right)^{1/6} \tag{6a}$$

$$\frac{dM_1}{dt} = 0 \tag{6b}$$

$$\frac{dM_2}{dt} = -\frac{\sqrt{2} B_1 (701 M_C^2 - 4210 M_C - 6859) M_1^2}{5184} \left(\frac{M_2}{M_1}\right)^{1/6} \tag{6c}$$

For collision frequency in the continuum regime, the corresponding moment equations are governed by

$$\frac{dM_0}{dt} = \frac{B_2 (2 M_C^2 - 13 M_C - 151) M_0^2}{81} \tag{7a}$$

$$\frac{dM_1}{dt} = 0 \tag{7b}$$

$$\frac{dM_2}{dt} = -\frac{B_2 (2 M_C^2 - 13 M_C - 151) M_1^2}{81} \tag{7c}$$

And the particle dimensionless moment is defined as

$$M_C = \frac{M_0 M_2}{M_1^2} \tag{8}$$

For Brownian coagulation, the scaling asymptotic growth rate can be found as (Xie & Wang, 2013):

$$-\frac{1}{M_0}\frac{dM_0}{dt} = \frac{1}{M_2}\frac{dM_2}{dt} = \frac{C}{t} \quad \text{and} \quad \frac{dM_C}{dt} = 0 \tag{9}$$

In which $C$ is a constant, and $C = 6/5$ in the free molecule regime, $C = 1$ in the continuum regime, respectively.

**The information entropy of SCE based on TEMOM**

Smoluchowski coagulation equation is deferent from the Boltzmann equation (Cercignani, 1988). In the former, the collision between particles is perfectly inelastic, the mass and momentum are conserved while the maximum amount of kinetic energy of system is lost, and the total particle number density is reduced. Therefore, Smoluchowski coagulation equation is a soft spheres collision model. However, the Boltzmann equation is a hard spheres collision model, the conservation of mass and momentum together with the conservation of kinetic energy makes possible the calculation of final velocities in two-body collision, and the particle number density remains constant. Therefore, the definition of information entropy for SCE is deferent from that for Boltzmann equation. In the latter, the information entropy is the only the function of the integral of the product of the distribution function and its logarithm.

In the literatures, there are several definitions of information entropy for SCE, such as the definition based on the coagulation probability (Gmachowski, 2001), but the effect of particle number density on the entropy is not considered. Another way to define the entropy for SCE is based on thermodynamics (Xie, 2017; Xie & Yu, 2018), the properties and evolution of entropy for SCE are studied from the perspective of energy balance and conversion, but not from the SCE itself.

Starting from the original definition of information entropy (Jaynes, 1957), which is a function of the total number of microscopic states ($\Omega$) in a disperse system, i.e., $S = k_B \ln \Omega$, and $\Omega$ can be calculated for SCE as

$$\Omega = \frac{M_0!}{\sum n!} \qquad (10)$$

According to the Sterling formula, then the information entropy can be expressed as

$$S = k_B[(M_0 \ln M_0 - M_0) - \sum(n \ln n - n)] \qquad (11)$$

Therefore, the information entropy for SCE can be considered as a function of PSD and the total number of particles.

Let the normalized PSD as

$$p(v,t) = \frac{n(v,t)}{M_0} \qquad (11)$$

Then the information entropy can be simplified as the form of continuity function

$$S = -k_B M_0 \int_0^\infty p(v,t) \ln p(v,t) \, dv \qquad (12)$$

**Principle of maximum entropy**

In mathematics, the principle of maximum entropy can be expressed as that the normalized PSD makes the information entropy approach to the extremum, i.e.,

$$\max[S(M_0, p(v,t)) = -k_B M_0 \int_0^\infty p(v,t) \ln p(v,t) \, dv] \qquad (13)$$

with the moment constraints as

$$\int_0^\infty p(v,t) dv = 1 \qquad (14a)$$

$$\int_0^\infty v p(v,t) dv = \frac{M_1}{M_0} \qquad (14b)$$

$$\int_0^\infty v^2 p(v,t) dv = \frac{M_2}{M_0} \qquad (14c)$$

Based on the variational principle and Lagrange multiplier method, the normalized PSD under the condition of maximum entropy can be obtained as a lognormal function:

$$p(v,t) = \frac{1}{3\sqrt{2\pi}\ln\sigma}\exp\left[-\frac{\ln^2(v/v_g)}{18\ln^2\sigma}\right]\frac{1}{v} \tag{15}$$

In which the geometric mean volume $(v_g)$ and standard deviation $(\sigma)$ in the distribution function can be expressed as the functions of particles moments:

$$v_g = \frac{M_1^2}{M_0^{3/2}M_2^{1/2}} \tag{16}$$

$$\ln^2\sigma = \frac{1}{9}\ln M_C \tag{17}$$

Then the information entropy at equilibrium can be expressed as

$$S^{eq} = k_B M_0\left[\frac{1}{2} + \frac{1}{2}\ln\frac{2\pi\ln M_C}{M_C} + \ln\frac{M_1}{M_0}\right] \tag{18}$$

Which is the function of particle dimensionless moment $M_C$ and zeroth order moment $M_0$. Using the extreme condition again, i.e.,

$$\frac{\partial S^{eq}}{\partial M_C} = 0 \text{ and } \frac{\partial S^{eq}}{\partial M_0} = 0 \tag{19}$$

The parameters of PSD at equilibrium can be determined as

$$M_C^{eq} = e, \ M_0^{eq} = M_1\sqrt{2\pi}/e \text{ or } v_g = \sqrt{e/2\pi}, \ \sigma = \sqrt[3]{e} \tag{20}$$

And the information entropy at equilibrium can be simplified as $S^{eq} = k_B M_0^{eq}$. The corresponding algebraic mean volume $(M_1/M_0)^{eq} = e/\sqrt{2\pi}\sim 1.0844$. The result reveals that the assumption that $(M_1/M_0)^{eq} = 1$ in the self-preserving hypothesis is reasonable in some sense.

**Cercignani's conjecture for SCE**

Cercignani's conjecture is based on the entropy-entropy production method, which was first used in kinetic theory for the Fokker-Planck equation (Toscani, 1999), it has been at the core of the renewal of the mathematical theory of convergence to thermodynamically equilibrium for Boltzmann equation over the past decade (Desvillettes & Villani, 2005). In explicitly, Cercignani's conjecture assumes a linear inequality between the entropy and the entropy production functional for Boltzmann integral operator as

$$\frac{dS}{dt} \geq K(S^{eq} - S) \tag{21}$$

In which $K$ is the proportional coefficient. Recently, Cercignani's conjecture for Boltzmann equation with hard sphere collision has been proved by Villani (2003). Brownian coagulation is a similar dissipative system as that described by Boltzmann equation. Is this conjecture still true for the SCE?

In 2019, Xie and Liu (2019) have proposed a similar equality of the relationship between entropy and entropy production for SCE as the structure of Cercignani's conjecture according to the thermodynamic constraint. In their derivation process, a similar $H$ function as that in Boltzmann equation is defined, but the relationship between the new $H$ function and entropy $S$ remains unknown. Because the entropy in that work is based on thermodynamics, it is difficult to answer this question theoretically. Can the definition of information entropy in present work solve this problem?

The entropy production can be calculated from the definition of entropy in Eq.(12) as

$$\frac{dS}{dt} = -\frac{1}{M_0}\frac{dM_0}{dt}\left[-M_0^2 \frac{d(S/M_0)}{dM_0} - S\right] \quad (22)$$

According to the principle of maximum entropy, the entropy is less than that at equilibrium, i.e., $S \leq S^{eq}$. In this way, the following inequality will hold,

$$\frac{dS}{dt} \geq -\frac{1}{M_0}\frac{dM_0}{dt}\left[-M_0^2 \frac{d(S^{eq}/M_0)}{dM_0} - S\right] = K(S^{eq} - S) \quad (23)$$

In which the constant $K$ is

$$K = -\frac{1}{M_0}\frac{dM_0}{dt} \quad (24)$$

Therefore, the Cercignani's conjecture is true for SCE due to Brownian motion, which means that the effect of molecular collisions is to force a non-equilibrium distribution function at a point in physical space back to a lognormal distribution function, and SCE for Brownian motion is convergence in mathematically.

**Discussions and conclusions**

In this study, the information entropy for SCE is proposed from the point of view of total number of microscopic states in a disperse systems. Based on the variational principle and Lagrange multiplier method, the normalized PSD can be obtained as a lognormal function under the condition of maximum entropy and moment constraints. The information entropy at equilibrium are the functions of particle dimensionless moment and zeroth order moment. Using the extreme condition, the geometric mean volume and standard deviation are both determined as simple constants. The results are consistent with the self-preserving hypothesis. It should be pointed out that the geometric mean volume has not been determined in the previous literatures.

With the present definition of information entropy, the Cercignani's conjecture for SCE holds naturally. The result reveals that the SCE for Brownian motion is convergence in mathematically. As mentioned above, Cercignani's conjecture are both true for elastic and perfectly inelastic collision theory. SCE is a soft sphere collision model, and Boltzmann equation is a hard sphere collision model. Moreover, a general collision can be decomposed into the weighted sum of elastic and perfectly inelastic collision. Therefore, Cercignani's conjecture holds for any two-body collision systems, which will explain many unknown problems. For example, the Navier-Stokes equation can be derived from the Boltzmann equation using Chapman-Enskog asymptotic expansion method, meanwhile, there is a dissipation of turbulent kinetic energy in turbulence flow. The question is that how can the dissipation of turbulent kinetic energy be realized in elastic collision system? Perhaps the combination of elastic and perfectly inelastic collision theory can solve this theoretical inconsistent problem.

**Acknowledgements**

This work is supported by the National Natural Science Foundation of China with Grant nos. 11972169 and 11572138.